# Molecular Simulation of Fracture Dynamics of Symmetric Tilt Grain Boundaries in Graphene

*Young In Jhon,[1] Pil Seung Chung,[2] Robert Smith,[2] and Myung S. Jhon[1,2]\**

[1]School of Advanced Materials Science and Engineering, Sungkyunkwan University, Suwon 440-746, Korea, [2]Department of Chemical Engineering and Data Storage Systems Center, Carnegie Mellon University, Pittsburgh, PA 15213, USA

**Abstract**

Atomistic simulations were utilized to obtain microscopic information of the elongation process in graphene sheets consisting of various embedded symmetric tilt grain boundaries (GBs). In contrast to pristine graphene, these GBs fractured in an extraordinary pattern under transverse uniaxial elongation in all but the largest misorientation angle case, which exhibited intermittent crack propagation and formed many stringy residual connections after quasi mechanical failure. The strings known as monoatomic carbon chains (MACCs), whose importance was recently highlighted, gradually extended to a maximum of a few nanometers as the elongation proceeded. These features, which critically affect the tensile stress and the shape of stress-strain curve, were observed in both armchair and zigzag-oriented symmetric tilt GBs. However, there exist remarkable differences in the population density and the achievable length of MACCs appearing after quasi mechanical failure which were higher in the zigzag-oriented GBs. In addition, the maximum stress and ultimate strain for armchair-oriented GBs were significantly greater than those of zigzag-oriented GBs in case of the largest misorientation angle while they were slightly smaller in other cases. The maximum stress was larger as the misorientation angle increased for both armchair and zigzag-oriented GBs ranging between 32~80 GPa, and the ultimate strains were between 0.06~0.11, the lower limit of which agrees very well with the experimental value of threshold strain beyond which mechanical failure often occurred in polycrystalline graphene.

---
\* *Corresponding author.* Fax: + 412-268-7139. Email address: mj3a@andrew.cmu.edu (M.S. Jhon)

# 1. Introduction

Graphene, a monolayer of hexagonally $sp^2$-bonded carbon atoms has been attracting great interest because of its elegant properties, such as ultrahigh electronic mobility [1,2], superior thermal conductivity [3,4], and excellent mechanical strength including exceptional stretchability [5-7]. These extraordinary properties in the defect free network can be obtained easily due to high formation energies of defects and strong bonding of carbon atoms [8]. Being much closer to commercial application, breakthroughs have recently occurred in the large-scale synthesis of the graphene film based on the chemical vapor deposition (CVD) technique [9,10]. This large-scale metallic catalyst based process, however, has intrinsically generated a polycrystalline form of graphene, owing to crystal imperfections of substrate material and kinetic influence in the growth process [11-14], and as a consequence, grain boundaries (GBs) have become one of the most dominating intrinsic defects. Graphene is unique in that it can host lattice defects through the reconstruction of atomic arrangement by forming non-hexagonal rings without inducing any broken bonds. For example, Stone Thrower Wales (STW) defects [15,16] do not involve any removed or added atoms, whereas single vacancy [17,18] undergoes Jahn-Teller distortion leading to the saturation of two of the three dangling bonds toward the missing atom. The zero-dimensional 5-7 defects in graphene can be aligned or merged to form a line which generates structurally robust GBs [19]. Considerable efforts have been paid to characterize the GBs in graphene [20-26], however their influence on the mechanical properties of graphene has rarely been reported due to great technological challenges.

To address this lingering issue, we investigated the mechanical behavior of polycrystalline graphene consisting of various symmetric tilt GBs using molecular dynamics (MD) simulations under uniaxial elongation in a direction transverse to GBs. In this paper we examined two classes of tilt GBs, namely, armchair-oriented (AC) and zigzag-oriented (ZZ) tilt GBs, with their structures depicted in Fig. 1. For each type of GBs, we examined three cases by decreasing misorientation angle, and they were denoted by $T_i$ (i=1~3, indicated by the descending order of the misorientation angle). We introduced symbols of $ACT_i$ and $ZZT_i$ to denote the sub-groups of armchair and zigzag-oriented tilt GB systems, respectively. The defect-free pristine (PR) graphene was also examined for comparison that was further subdivided into ZZPR or ACPR depending on the elongation direction though their structures are the same.

## 2. Computational Methods

All the $T_i$ systems were constructed to meet the periodic boundary condition by embedding two directionally opposite tilted GBs to a pristine graphene. The MD simulations were performed using the software package LAMMPS [27] with the adaptive intermolecular reactive empirical bond order (AIREBO) potential [28] and time step of 1.0 fs. The periodic dimensions of the simulation system and atomic coordinates were first optimized using a gradient-based minimization method with tolerance criteria of $10^{-8}$ eV/ Å in force and/or $10^{-8}$ eV in energy. Based on the simulation cell size obtained from the above calculation, NVT simulation was performed consecutively for 300 000 steps, and finally the system was elongated uniaxially in the direction perpendicular to GBs. The strain rate was set to 0.1 ns$^{-1}$ and 2 000 steps were taken at each deformation point, where 0.02% strain was applied to the system between two consecutive points. Non-equilibrium molecular dynamics (NEMD) simulations of a continuously strained system were performed using SLLOD equations of motion coupled to a Nose-Hoover thermostat [29].

In the AIREBO potential, the cut-off radius was set to be 2.0 Å to avoid spuriously high bond forces and unphysical results near the fracture region. This value has tacitly been used to study mechanical properties of graphene [6,30,31]. In an alternative approach, the value of 1.92 Å was used to obtain a better quantitative agreement with experimental data for hydrogenated graphene and generated successful results on graphene characteristics [32]. However, it causes spurious consequences in pristine graphene as shown in our previous study [26] where a serious discrepancy was observed between quantum mechanical calculation and classical force field calculation based on the AIREBO potential using cut-off radius of 1.92 Å in elongation behavior of $ZZT_2$ system, while they showed a good coincidence in the case using cut-off radius of 2.0 Å.

## 3. Results and discussion

Pristine graphene was elongated in zigzag and armchair directions and we observed that the stress and strain at tensile failure were larger for zigzag-directional elongation than for the armchair-direction as shown in Fig. 2 (a), which was in good agreement with previous studies [30,32]. Their magnitudes were, however, somewhat smaller than published values including those of our earlier study [26], which is due to much slower and refined strain change between two consecutive stages as indicated by Zhao et al.'s study of the strain rate effect [30]. A short damped oscillation was observed in stress after tensile failure for armchair-directional elongation which resulted from the temporary elastic motion of graphene after complete fracture. This was not observed when larger steps were taken in the averaging of stress at each deformation point. The fractured structures emerging at tensile failure are shown in Figs. 3 (a) and (b) for ZZPR and ACPR systems, respectively. These figures indicated that fracture always occurred along the zigzag direction, independent of tensile load direction. We believe this to be the reason for the larger magnitudes of ultimate strength and strain under zigzag-directional load, compared to those of the armchair-direction, since the normal direction of the fracture plane does not coincide with the tensile load direction in the former case. This leads to smaller effective tensile load to the fracture plane while such a phenomenon did not occur in the case of armchair-directional load.

The tensile stress-strain curves for graphene sheets with various tilt GBs under transverse uniaxial elongation are presented in Fig. 2. The curve pattern of $T_1$ systems is rather similar to pristine graphene. However, $T_2$ and $T_3$ systems exhibited a peculiar pattern, irrespective of whether they are armchair or zigzag-oriented. Their fractured structures at tensile failure are shown in Figs. 3 (c)-(f) where we observed the formation of dense population of monoatomic carbon chains (MACCs) connecting fracture sections, which recently become a headline in designing novel graphene micro- and nanoelectromechanical systems (M/NEMS) [33].

Two unique patterns in stress-strain curves, namely, a remarkable decrease in the slope (with rugged stress pattern) prior to tensile failure and subsequent saw-teeth shaped fluctuation pattern were observed for both armchair and zigzag-oriented tilt GBs. These features were also found in our precedent study for mechanical characteristics of graphene having zigzag-oriented tilt GBs where elongation rate was 0.5 ns$^{-1}$ [26]. The structural analysis indicates that those patterns were caused by intermittent crack propagation prior to tensile failure and subsequent incomplete mechanical failure producing long-lasting MACCs connecting two separated fractured sections. Tensile stress decreased considerably either when the MACCs were extended chemically using carbon atoms supplied from attached graphene through the bond-breaking/reforming process or when the string was disconnected permanently due to elongation. The same behavior was observed in armchair-oriented GB systems as shown in Figs. 4 (a)-(c) where the structural change of the ACT$_3$ system is illustrated. This peculiar mechanical behavior is also captured in our tentative study using a cut-off radius of 1.92 Å as shown in Fig. S1, although the stress fluctuation is lower than that of 2.0 Å.

Focusing on the elongation and fracture of symmetric tilt GBs, we performed in-depth analysis on the population density variation of MACCs during the elongation process as shown in Fig. 5. Among all examined cases, the longest and most dense MACCs were obtained in cases of T$_2$ and T$_3$ systems, where zigzag-oriented tilt GBs were especially preferable to armchair-oriented tilt GBs. The MACC formation in ZZT$_2$ system is illustrated in Fig. 4 (d).

From the complex stretching pattern, we define the tensile failure point as the time at which graphene begins to split into two fractured parts being separated completely or connected only by MACCs. According to this definition, strain at tensile failure was plotted as a function of misorientation angle for $ACT_i$ and $ZZT_i$ systems as shown in Fig. 6 (a). We found that it ranged between 0.06~0.11 for all cases, which supports the experimental result that large-scale graphene film transferred on an unstrained substrate recovered its original resistance after stretching of 6%, however, further stretching often results in mechanical failure [34].

The maximum stress was also plotted as a function of misoritentation angle, shown in Fig. 6 (b). The result showed that the maximum stress was larger as misorientation angle increased for both armchair and zigzag-oriented tilt GBs ranging 32~80 GPa, which indicates that GBs having higher density of defects sustained larger tensile stress. This counterintuitive trend was first observed in a previous study for graphene consisting of zigzag-oriented tilt GBs [24] and was understood by considering the critical bonds in the heptagonal carbon rings that lead to failure. These bonds were more extended initially as the misorientation angle decreased, being more susceptible to the fracture.

In our study, we found that, for various $ACT_i$ systems, the critical bonds are always located in the heptagonal ring being adjacent to nearby hexagon (or pentagon in case of $ACT_1$), similar to $ZZT_i$ systems. The averaged value of the initial lengths for the critical bonds was plotted as a function of misorientation angle as shown in Fig. 7. As expected, the bond length was shorter as misorientation angle increased in both $ACT_i$ and $ZZT_i$ systems, which indicates that an identical mechanism can be applied to $ACT_i$ systems to explain their abnormal tensile strengths opposite to the degree of defect density. The tensile strengths of $ACT_2$ and $ACT_3$ systems are smaller than those of $ZZT_2$ and $ZZT_3$ systems, respectively, while the strength of $ACT_1$ is much larger than that of $ZZT_1$ as shown in Fig. 6 (b), and this can be captured through the analysis on initial lengths of critical bonds as shown in Fig. 7, implying its capability to predict tensile strength accurately.

In contrast to the coincidence of the points for maximum stress and tensile failure in cases of the PR and $T_1$ systems, it is noteworthy that they did not occur simultaneously for $T_2$ and $T_3$ systems as indicated by circles and arrows respectively in Figs. 2 (a)-(d).

## 4. Conclusion

The existence of symmetric tilt GBs generated an extraordinary tensile behavior in all but the largest misorientation angle case under transverse uniaxial elongation. The stress-strain curve exhibits a prominent decrease in slope prior to tensile failure and subsequent long-standing saw-teeth shaped fluctuation pattern in stress-strain plot for both armchair and zigzag-oriented tilt GBs. Furthermore, our structural analysis for armchair-oriented tilt GBs verified the hypothesis that the magnitude of tensile stress is adversely influenced by the initial length of the critical bond that is the most susceptible to elongation. It was shown that despite these qualitative similarities in fracture behavior between armchair and zigzag-oriented GBs, there exists the significant anisotropic difference in key characteristics such as the maximum available length and population of MACCs, the importance of which was recently highlighted in graphene M/NEMS. The formation of MACCs results in the long standing stress patterns after tensile failure, especially for the zigzag-oriented graphene.

We believe that these findings will contribute greatly in understanding and developing the polycrystalline graphene based devices at a more fundamental level in the future.

## Acknowledgement

This work was supported by the World Class University program of KOSEF. (Grant No. R32-2008-000-10124-0).

**Figure captions**

Figure 1. The structures of grain boundaries (a)-(c) in zigzag-oriented graphene sheets ($ZZT_1$, $ZZT_2$, and $ZZT_3$ systems, respectively) and (d)-(f) in armchair-oriented graphene sheets ($ACT_1$, $ACT_2$, and $ACT_3$ systems, respectively).

Figure 2. The stress-strain curves for (a) PR, (b) $T_1$, (c) $T_2$, and (d) $T_3$ systems under uniaxial elongation. Circles and arrows indicate maximum stress and tensile failure, respectively.

Figure 3. The fracture structures at tensile failure for (a) ZZPR, (b) ACPR, (c) $ZZT_2$, (d) $ACT_2$, (e) $ZZT_3$, and (f) $ACT_3$ systems where MACCs are denoted in green.

Figure 4. The structures of $ACT_3$ system (a) at crack creation denoted by red ellipsoids, (b) at crack propagation corresponding to the precipitous drop in stress prior to tensile failure, (c) under strain of 0.1138 exhibiting elongated monoatomic carbon chains between two fractured sections, and (d) the structure of $ZZT_2$ system under strain of 0.15.

Figure 5. The duration and change in the population density of MACCs under elongation of graphene perpendicular to (a) armchair-oriented GBs and (b) zigzag-oriented GBs in conjunction with those of pristine graphene. Zigzag-oriented GBs produced longer and denser MACCs than armchair-oriented GBs for all $T_i$ systems and especially, $ZZT_2$ and $ZZT_3$.

Figure 6. (a) The tensile failure strain and (b) maximum stress as a function of misorientation angle for graphene having armchair and zigzag-oriented tilt GBs under transverse uniaxial elongation.

Figure 7. The plot of initial critical bond length as a function of misorientation angle for graphene having armchair and zigzag-oriented GBs.

Figure S1. The stress-strain plots of (a) $T_2$ and (b) $T_3$ systems under uniaxial elongation using

cut off radius of 1.92 Å .

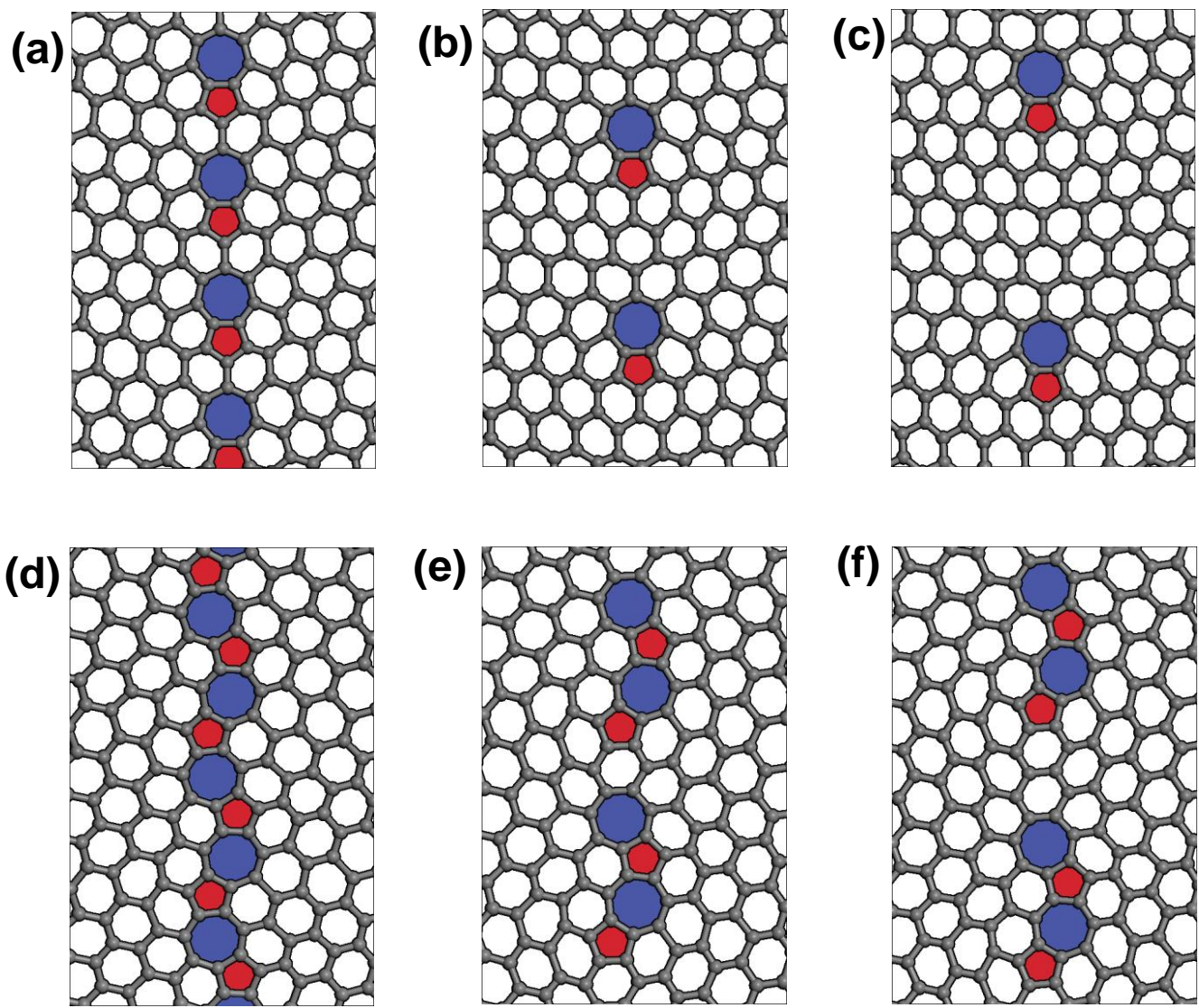

**Figure 1.** The structures of grain boundaries **(a)-(c)** in zigzag-oriented graphene sheets (ZZT$_1$, ZZT$_2$, and ZZT$_3$ systems, respectively) and **(d)-(f)** in armchair-oriented graphene sheets (ACT$_1$, ACT$_2$, and ACT$_3$ systems, respectively).

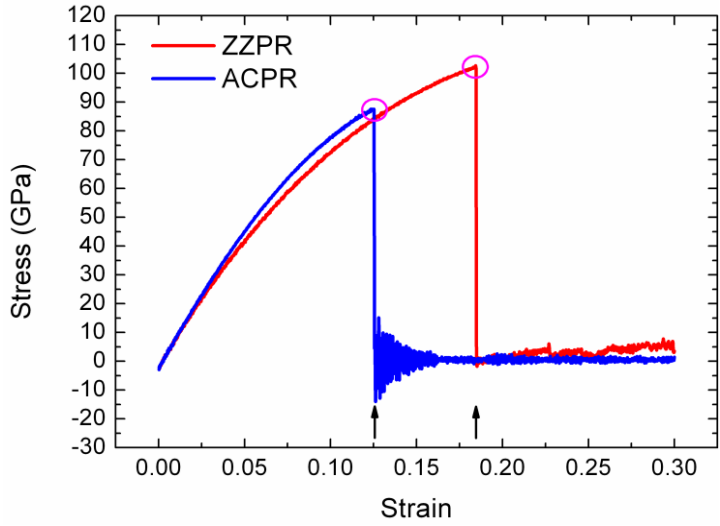
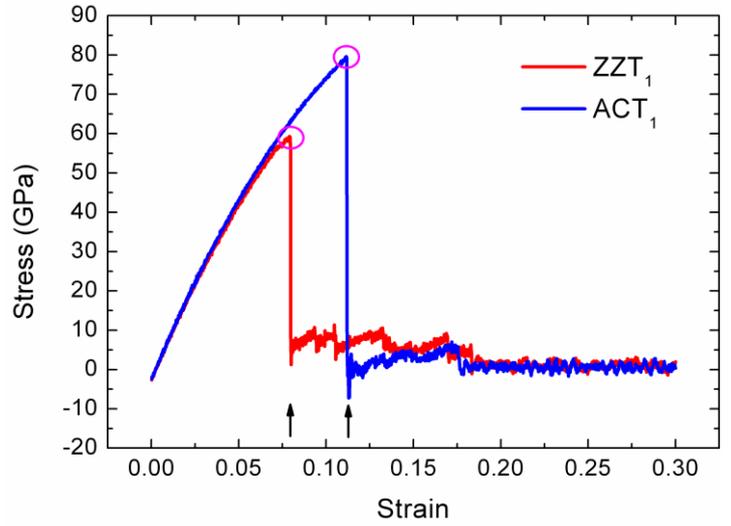
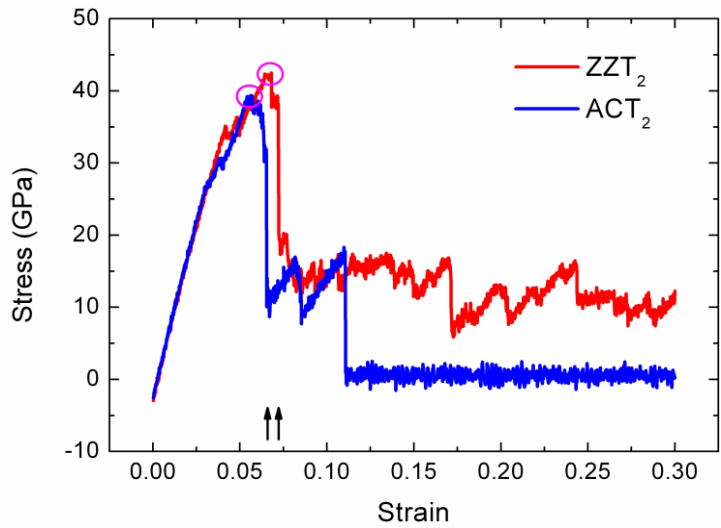
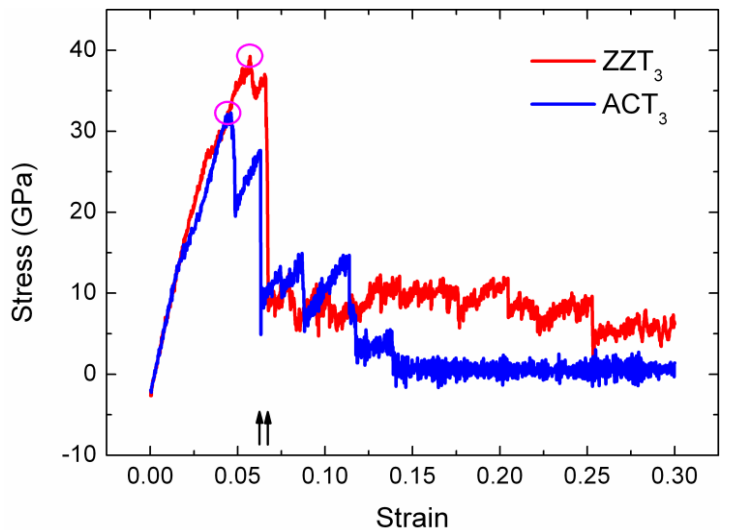

**Figure 2.** The stress-strain curves for **(a)** PR, **(b)** $T_1$, **(c)** $T_2$, and **(d)** $T_3$ systems under uniaxial elongation. Circles and arrows indicate maximum stress and tensile failure, respectively.

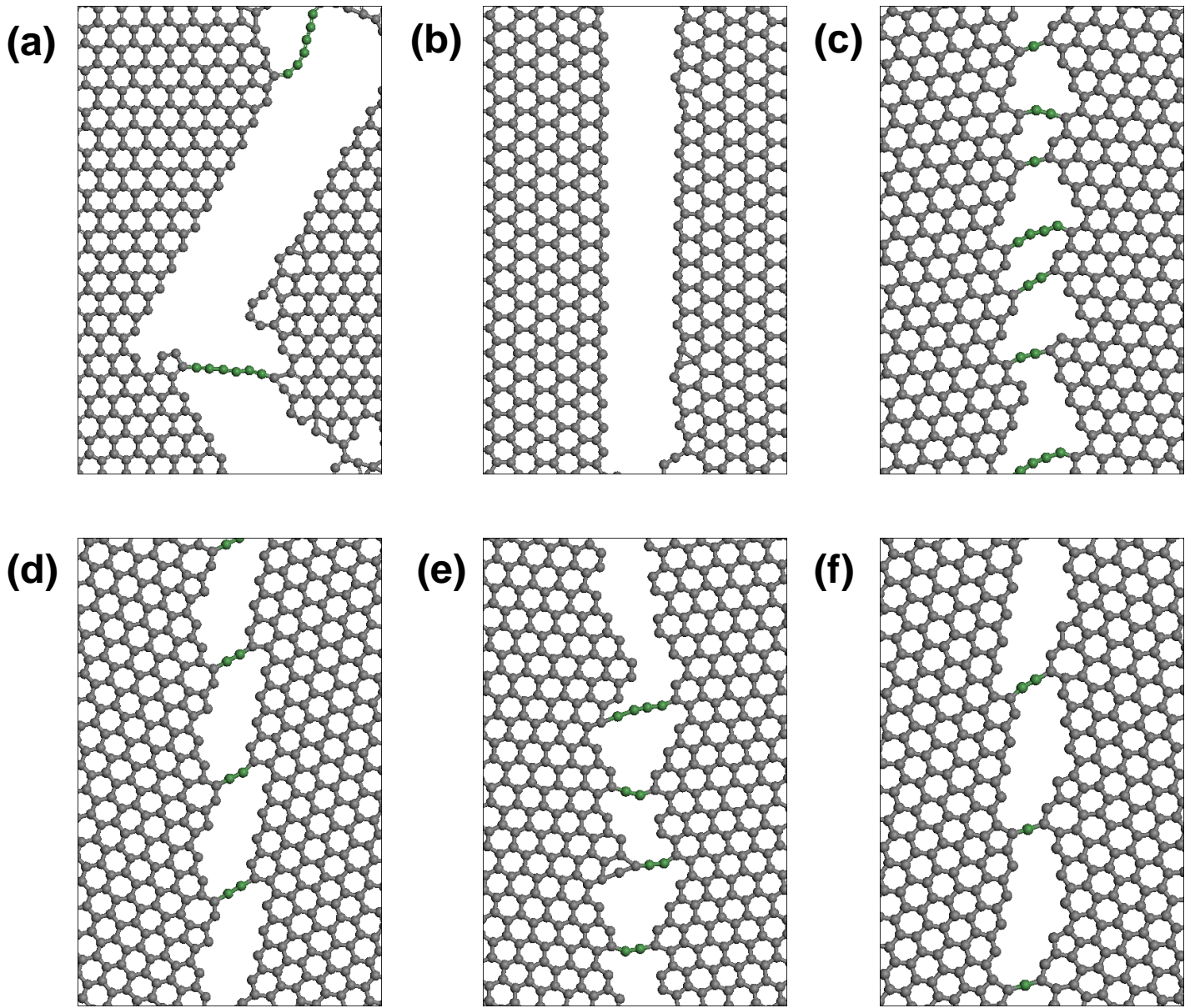

**Figure 3.** The fracture structures at tensile failure in **(a)** ZZPR, **(b)** ACPR, **(c)** $ZZT_2$, **(d)** $ACT_2$, **(e)** $ZZT_3$, and **(f)** $ACT_3$ systems where MACCs are denoted in green.

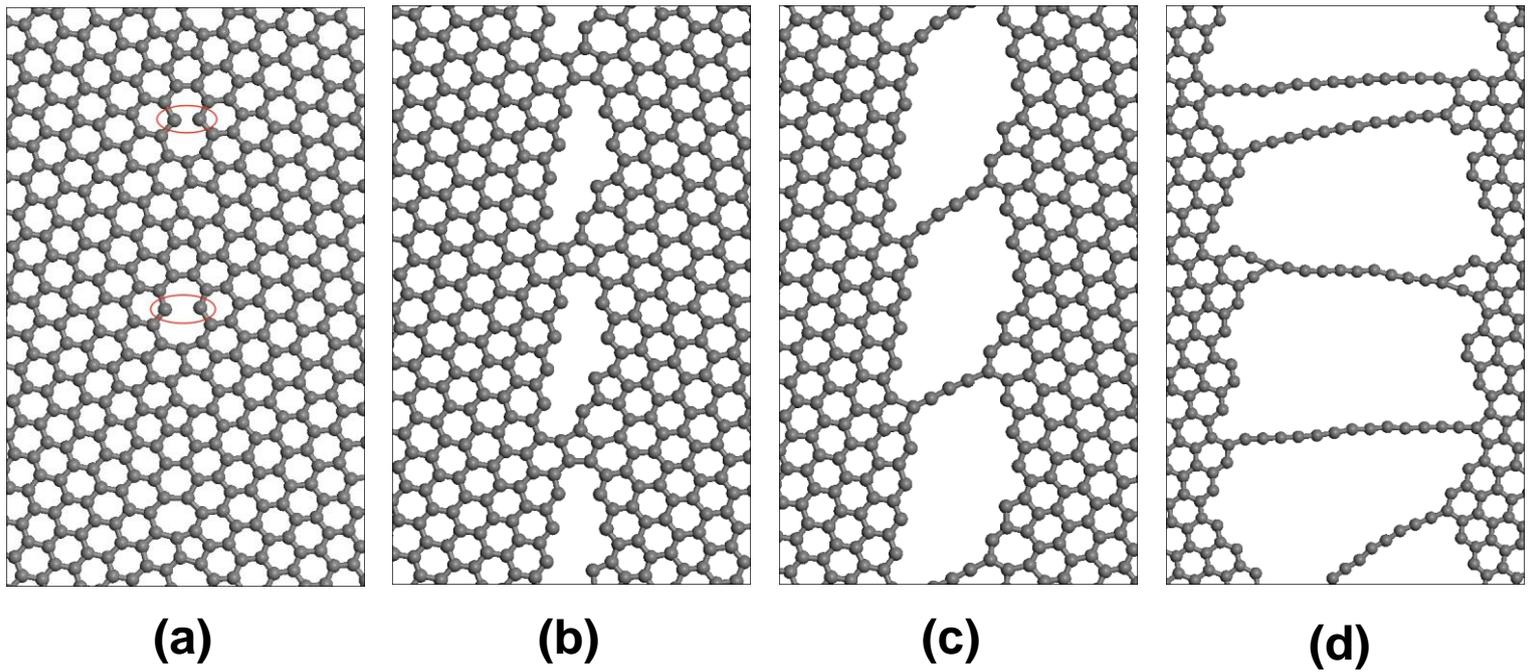

**Figure 4.** The structures of ACT$_3$ system **(a)** at crack creation denoted by red ellipsoids, **(b)** at crack propagation corresponding to the precipitous drop in stress prior to tensile failure, **(c)** under strain of 0.1138 exhibiting elongated monoatomic carbon chains between two fractured sections, and **(d)** the structure of ZZT$_2$ system under strain of 0.15.

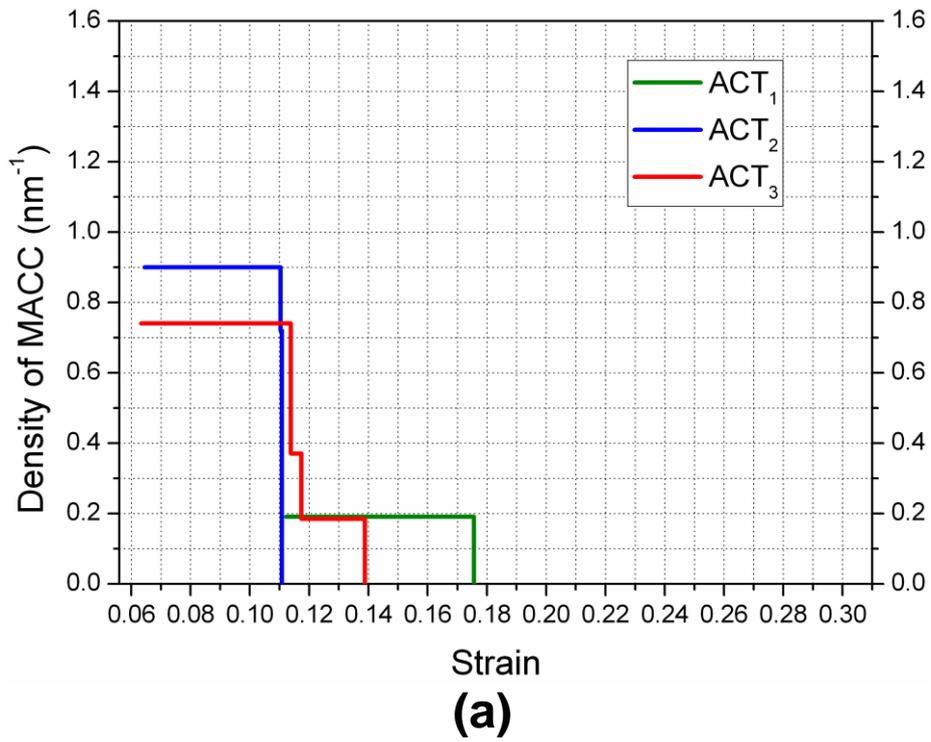

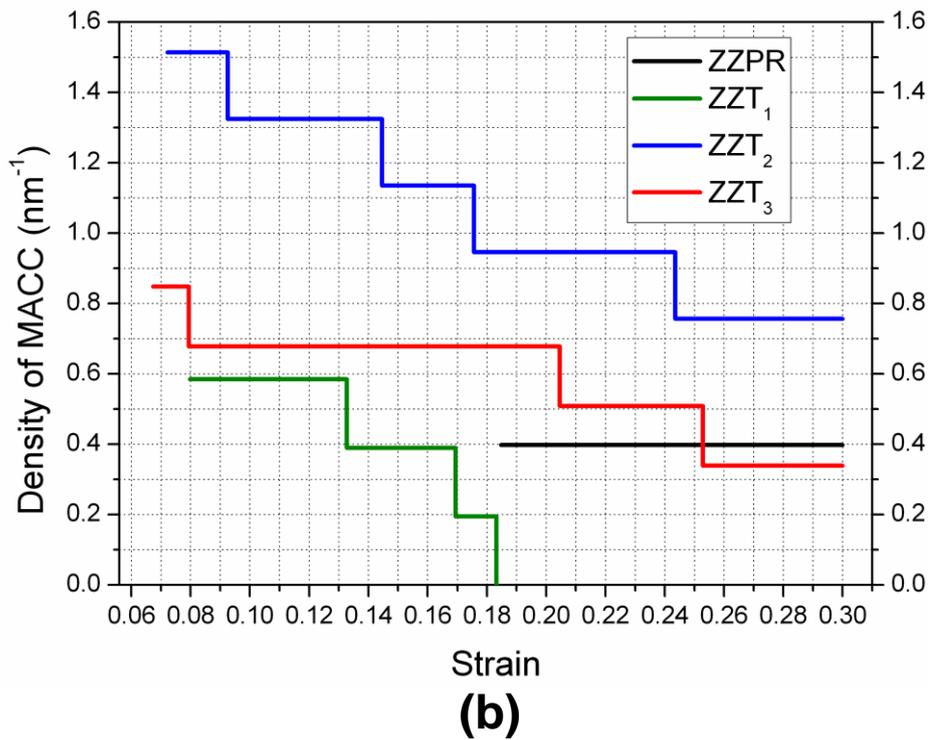

**Figure 5.** The duration and change in the population density of MACCs under elongation of graphene perpendicular to **(a)** armchair-oriented GBs and **(b)** zigzag-oriented GBs in conjunction with those of pristine graphene. Zigzag-oriented GB produced longer and denser MACCs than armchair-oriented GB for all $T_i$ systems and especially, $ZZT_2$ and $ZZT_3$.

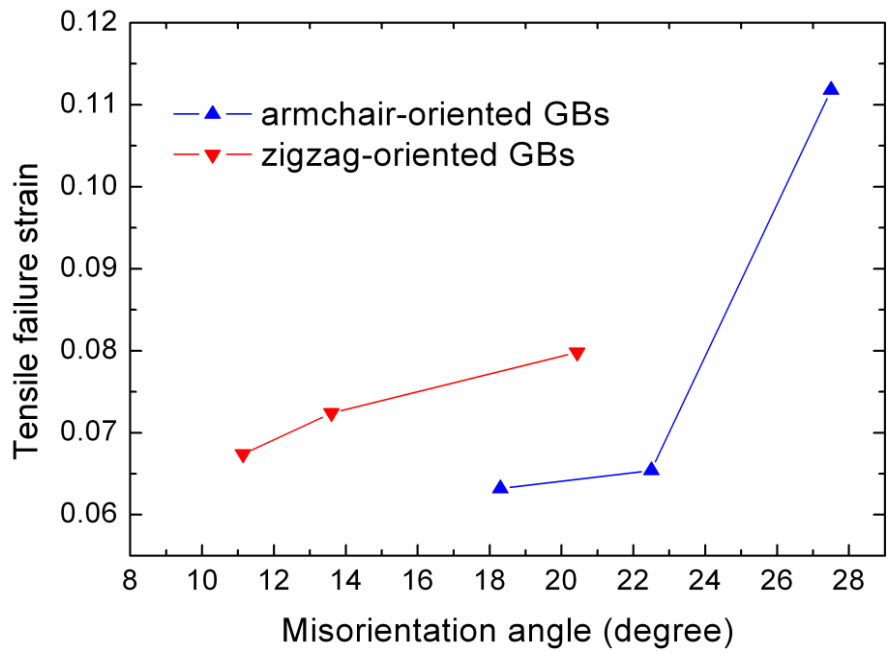

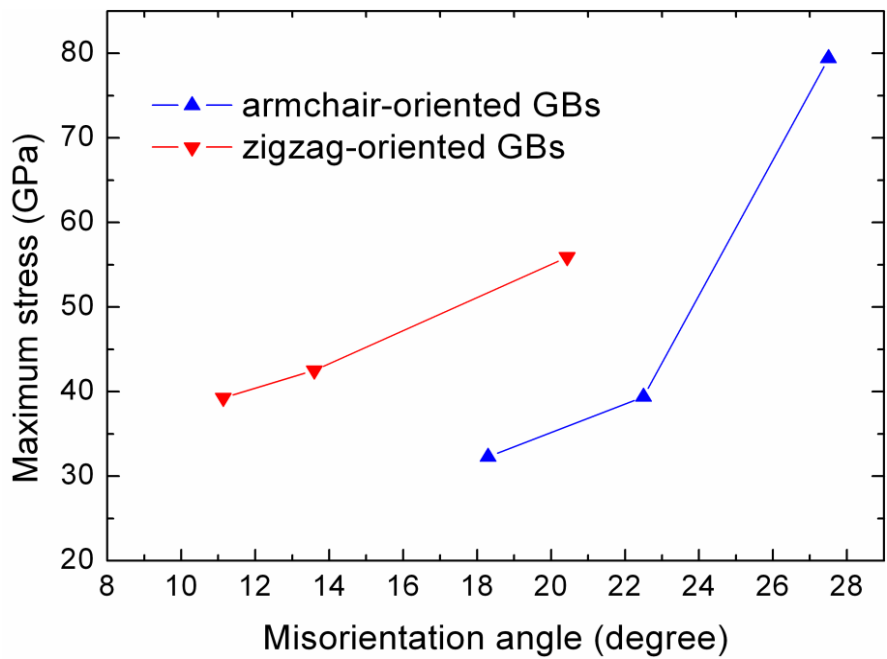

**Figure 6. (a)** The tensile failure strain and **(b)** maximum stress as a function of misorientation angle for graphene having armchair and zigzag-oriented tilt GBs under transverse uniaxial elongation.

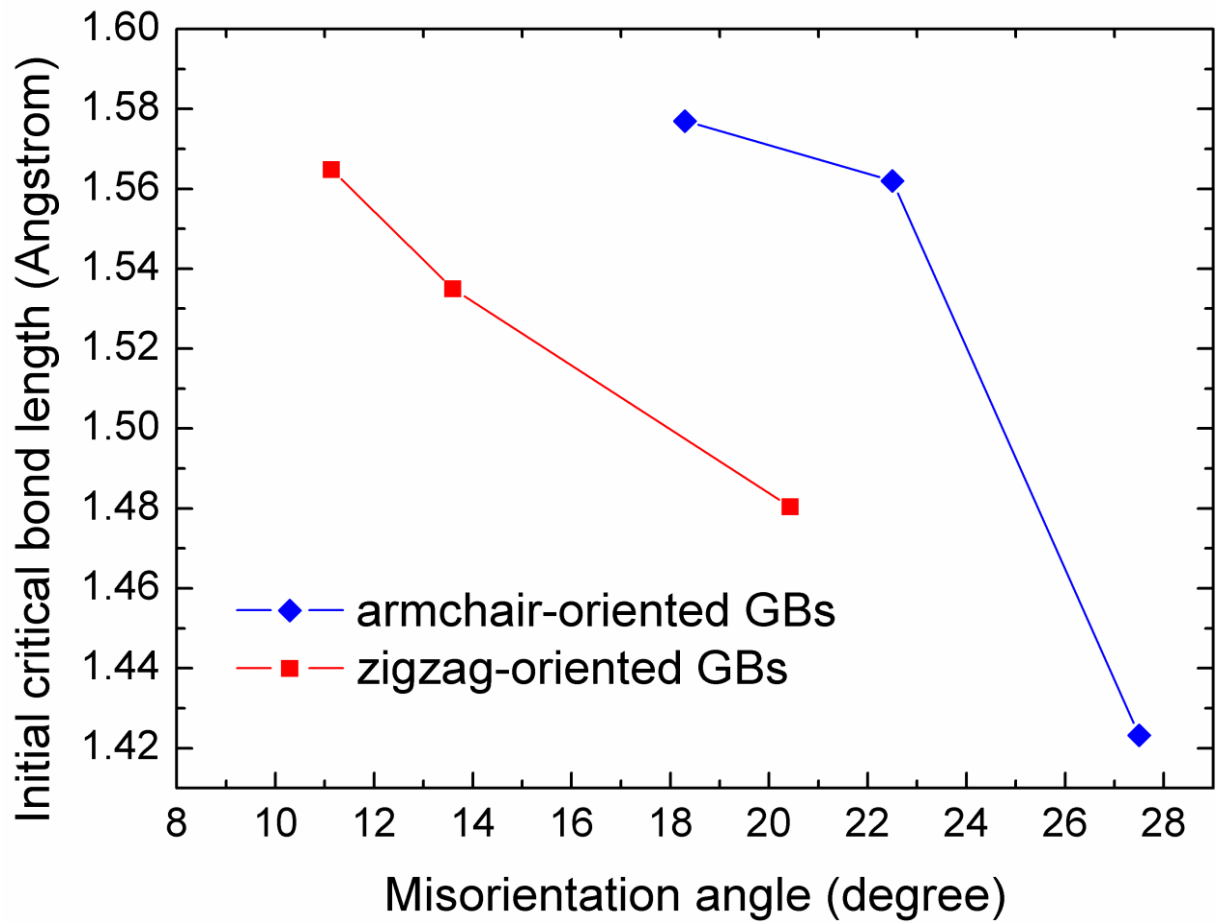

**Figure 7.** The plot of initial critical bond length as a function of misorientation angle for graphene having armchair and zigzag-oriented GBs.

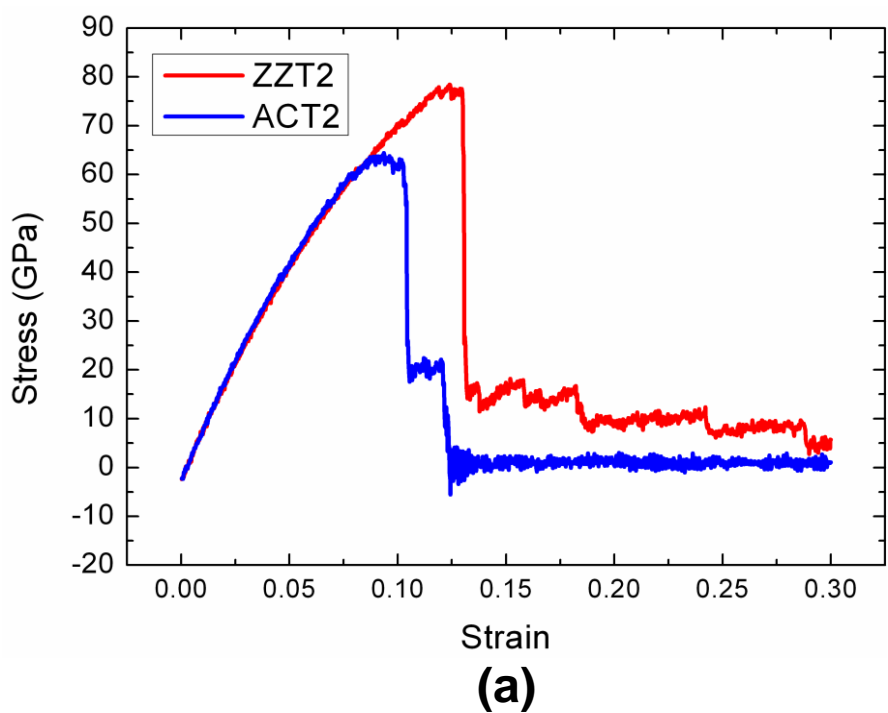

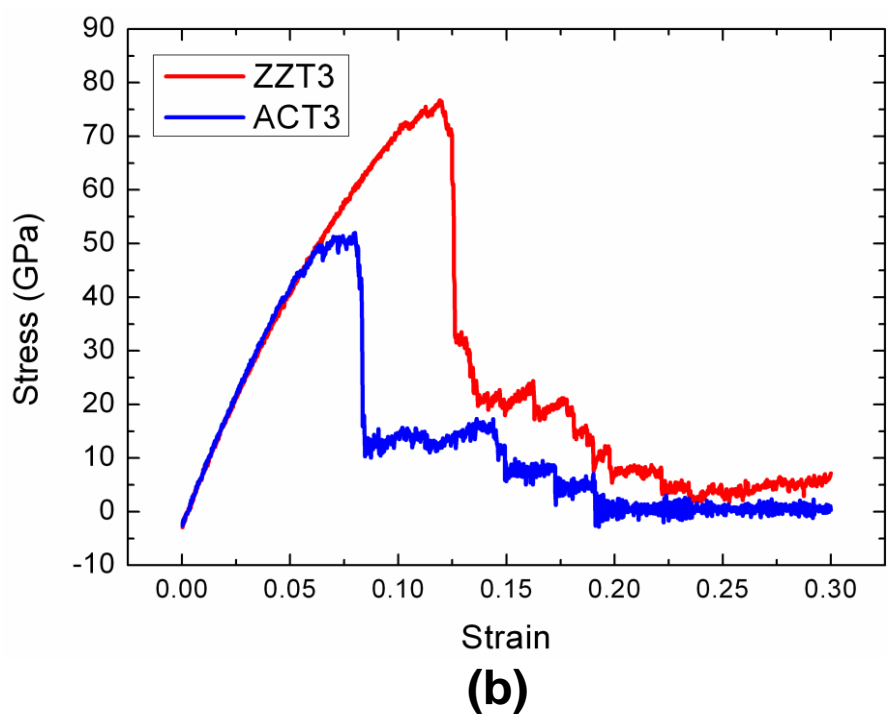

**Figure S1.** The stress-strain plots of **(a)** $T_2$ and **(b)** $T_3$ systems under uniaxial elongation using cut-off radius of 1.92 Å.